\documentclass[sigconf,nonacm]{acmart}

\AtBeginDocument{%
  }

\setcopyright{acmlicensed}
\copyrightyear{2018}
\acmYear{2018}
\acmDOI{XXXXXXX.XXXXXXX}
\acmConference[Conference acronym 'XX]{Make sure to enter the correct
  conference title from your rights confirmation email}{June 03--05,
  2018}{Woodstock, NY}
\acmISBN{978-1-4503-XXXX-X/2018/06}

\usepackage{hyperref}
\usepackage{cleveref} 
\usepackage{tabularx}
\usepackage{natbib}
\begin{document}

\title{RealityTrace: An Extended Reality Interface for Human-AI Embodied Problem Solving}

\author{Yunhao Luo}
\email{yunhaoluo@ucsb.edu}
\orcid{0009-0004-6219-8021}
\affiliation{%
  \department{Computer Science}
  \institution{University of California}
  \city{Santa Barbara}
  \state{CA}
  \country{USA}
}

\author{Arthur Caetano}
\email{caetano@ucsb.edu}
\orcid{0000-0003-0207-5471}
\affiliation{%
  \department{Computer Science}
  \institution{University of California}
  \city{Santa Barbara}
  \state{CA}
  \country{USA}
}

\author{Irene Li}
\email{irene_li@ucsb.edu}
\orcid{0009-0004-4806-7463}
\affiliation{%
  \department{Computer Science}
  \institution{University of California}
  \city{Santa Barbara}
  \state{CA}
  \country{USA}
}

\author{Tobias Höllerer}
\email{holl@cs.ucsb.edu}
\orcid{0000-0002-6240-0291}
\affiliation{%
  \department{Computer Science}
  \institution{University of California}
  \city{Santa Barbara}
  \state{CA}
  \country{USA}
}

\author{Misha Sra}
\email{sra@cs.ucsb.edu}
\orcid{0000-0001-8154-8518}
\affiliation{%
  \department{Computer Science}
  \institution{University of California}
  \city{Santa Barbara}
  \state{CA}
  \country{USA}
}

\renewcommand{\shortauthors}{Luo et al.}



\begin{abstract}
While XR-based task guidance has been widely explored through in-situ visualizations and step-by-step instructions, these systems assume predefined procedures and documented solutions. Many real-world physical problems lack this structure, requiring users to form hypotheses, test approaches, and backtrack through multiple possible solution paths. We present RealityTrace, an XR interface designed to support process documenting and tracing during open-ended, embodied problem solving. RealityTrace helps users externalize observations, track hypotheses, and record intermediate actions directly within the physical workspace. This supports tasks such as hardware debugging, device repair, and multi-path puzzles, where neither the user nor the system has prior knowledge of the solution. By capturing the embodied problem-solving process as it unfolds, RealityTrace enables users to revisit previously attempted procedures, organize alternative approaches, and share documented solution paths and outcomes for collaborative problem solving.
\end{abstract}




\begin{teaserfigure}
  \includegraphics[width=\textwidth]{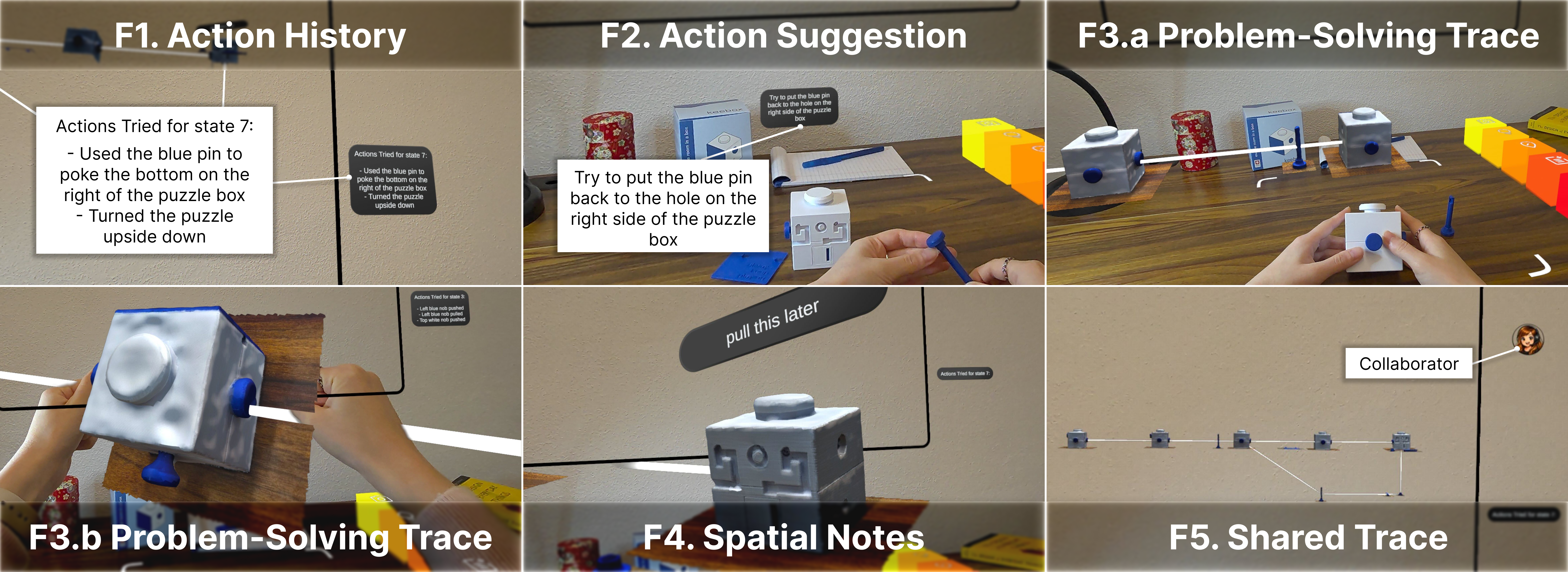}
  \caption{We present RealityTrace, an interface that enables real-time documentation and tracing in human-AI embodied problem solving: \textbf{F1. Action History} documents all previously attempted actions in each problem-solving state to prevent unproductive action loops. \textbf{F2. Action Suggestion} uses action history to suggest previously unexplored strategies to mitigate unproductive approaches. \textbf{F3.a Problem-Solving Trace} creates a persistent external representation of the problem-solving process and preserves a visual history of previously explored strategies, and \textbf{F3.b Problem-Solving Trace} allows users to manipulate the 3D reconstructed models and view states from different angles. \textbf{F4. Spatial Notes} enables users to attach persistent notes in the problem-solving trace or physical environment. \textbf{F5. Shared Trace} enables collaboration between multiple users by sharing and merging their problem-solving traces.}
  \Description{The figure presents six screenshots highlighting the five features of RealityTrace, an interface that enables real-time documentation and tracing in human-AI embodied problem solving. The first screenshot captures F1. Action History, the feature that documents all the actions the user has attempted on that certain stage. The second screenshot captures F2. Action Suggestion, the feature that provides hints to guide the user during their problem-solving process. The third screenshot F3.a Problem-Solving Trace and fourth screenshot F3.b Problem-Solving Trace capture the feature that allows users to create a persistent external representation (3D reconstruction) of the problem-solving process. In the fourth screenshot F3.b, the user is able to manipulate the 3D reconstructed models and view the states from different angles. The fifth screenshot F4. Spatial Notes captures the feature that allows users to annotate either on the 3D reconstructed model or the real world. The sixth screenshot F5. Shared Trace highlights the feature that allows users to collaborate and share their progress on the problem-solving process with others.}
  \label{fig:teaser}
\end{teaserfigure}

\maketitle

\section{Introduction}

In contrast to purely abstract problem solving, embodied problem solving, such as diagnosing and repairing a malfunctioning car, involves bodily actions that actively contribute to finding solutions, rather than merely executing predefined plans or decisions formed solely from internal reasoning~\cite{werner2019moving, spiridonov2019role}. This role of bodily action is particularly relevant in novel, open-ended problem scenarios, where solutions cannot be derived through established procedures and instead require active exploration and experimentation. Such processes often involve \textit{epistemic actions} that reveal information and facilitate reasoning, alongside \textit{pragmatic actions} that directly advance toward a solution~\cite{kirsh1994epistemic_pragmatic, esteves2015atb}. For instance, when troubleshooting an unexplained engine rattle, a mechanic might rev the engine to help identify the source of the noise (epistemic) before replacing a faulty chain to resolve the issue (pragmatic). As exploration progresses, successive epistemic actions can reveal an increasing amount of information that must be integrated with prior observations to inform subsequent decisions. Keeping track of what has been tried, what has been learned, and how the physical environment has changed can become cognitively demanding, particularly in open-ended tasks that require extensive exploration, where it is often unclear whether one's actions are leading toward a solution or how much progress has been made.

External representations can alleviate these demands by keeping relevant information about the problem readily accessible through an external medium, separate from both the object involved in the problem and the problem-solver's mind, thereby reducing the need to maintain this information in working memory~\cite{zhang1994representations}. Beyond reducing memory demands, external representations can also shape how people reason about physical problems by influencing what information is available and how is it interpreted~\cite{KIRSH199531}. For example, a mechanic might record engine sounds and annotate an engine diagram to connect observed symptoms with potential causes. Such representations can help reveal relationships or possibilities that might otherwise go unnoticed, informing further exploration and guiding subsequent actions~\cite{kirsh2010representations, zhang1997nature, larkin1987diagram}.

However, existing approaches to externalizing embodied problem solving present two key challenges. First, manually creating structured representations, such as annotated sketches or notes, requires users to divert their attention and physical actions away from the problem at hand toward a separate documentation activity, such as picking up a pen and writing on paper~\cite{tatzgern2010compact, kalkofen2009explosion}. Second, low-effort alternatives, such as video recordings, can capture the unfolding process with minimal interruption but often contain substantial amounts of information with varying relevance, without explicit structures for organizing it~\cite{yang2023beyond, kim2014crowdsourcing}. Although such recordings preserve rich details about actions and their context, extracting useful insights requires users to retrospectively search through and interpret the recorded material. Consequently, problem-solvers face a trade-off between structured representations that are effortful to create and rich recordings that are difficult to navigate and reuse. This motivates tools that capture and structure problem-relevant information as it emerges, allowing external representations to develop alongside the problem-solving process rather than requiring a separate, attention-demanding documentation activity.

To address the current lack of such a tool, we present RealityTrace, an extended reality interface that enables real-time creation of structured external representations during embodied problem solving. RealityTrace is designed for problem solving in discovery contexts where solutions are unknown and structured representations do not yet exist. 
RealityTrace enables users to document the problem-solving process as a task graph (trace) where each node represents a problem state and edges represent pragmatic actions that transform one state into another. Nodes contain multimodal data, including text observations, images, and 3D reconstructions that users can manipulate in AR to inspect problem states from different perspectives. Edges connect states and encode alternative solution paths as branches in the graph. The temporal structure of the trace enables users to track what has been tried, identify dead ends, and build on prior discoveries. This evolving structured representation both informs AI assistants that offer real-time suggestions and enables collaborative problem solving through shared traces. Critically, users retain agency over when to capture states, allowing them to decide how to structure the trace while offloading the technical work of 3D generation to the system.

RealityTrace's design was informed by a formative study and grounded in problem-solving literature~\cite{sweller1988cognitive, kirsh1994epistemic_pragmatic, bilalic2008einstellung}. We conducted an initial evaluation of RealityTrace through a usability study (n=8) using a Wizard-of-Oz backend, accessing usability with the system usability scale (SUS) and analyzing semi-structured interviews through reflexive thematic analysis. We contribute: \textbf{(1)} RealityTrace, an XR interface that enables efficient, real-time capture of structured representations during embodied problem solving; \textbf{(2)} empirical findings with implications for the design of embodied problem-solving tools. Our results demonstrate the potential of structured external representations for novel embodied problems and their value in scaffolding embodied problem solving and distributing it across AI agents and humans while preserving user agency.

\section{Related Work}

This work builds upon prior research on XR-based systems for computer-assisted embodied problem solving.

\subsection{Task Assistance in XR}

For problems with established solutions, an extensive body of work has explored task guidance using AR systems~\cite{WANG2022102407, cao2020exploratory, huang2021adaptutar, cao2022mobiletutar, liu2023instrumentar}. These systems use situated visualization to guide users through pre-curated procedures, transferring expert knowledge to end users while reducing task time and error rates~\cite{yang2019influences}. Authoring such guidance often relies on structured representations of tasks, such as action graphs created manually or derived from demonstrations~\cite{shi2025caring, patel2026assemblifyvr}. In industrial contexts, remote assistance platforms (e.g., Vuforia Chalk~\cite{ptc_remote_assistance}, Capgemini Andy 3D~\cite{capgemini_andy3d}) similarly support task execution by enabling experts to guide users through predefined or collaboratively constructed procedures. While effective, these approaches typically require substantial pre-authoring or prior demonstrations~\cite{nebeling2018trouble}, limiting their scalability and adaptability to unseen scenarios.

To address this limitation, recent work has explored the ad-hoc generation of AR guidance using large language models (LLMs)~\cite{zhao2025guided, stover2024taggar, deepmind2024astra}. These approaches draw on knowledge encoded in LLMs, as well as external information sources, to dynamically generate task-specific guidance procedures. Although they offer greater generalizability, they remain constrained by the scope, specificity, and reliability of previously documented knowledge incorporated into training data and its applicability to the unique problem at hand that does not have directly transferable solution procedures~\cite{zhang2024out}.

More fundamentally, many embodied problem-solving tasks require iterative exploration to identify underlying issues and their causes, rather than following predefined procedures. RealityTrace positions the user as the central agent in the problem-solving process, supporting and scaffolding their cognitive processes while balancing the effort and utility of external representations in problem solving.

\subsection{Sensemaking in Open-Ended Problems}

In open-ended tasks, sensemaking is a critical, iterative process through which problem solvers progressively develop an understanding of the task, often supported by the use of external tools. Prior work has explored ways to support this process by capturing and organizing users' exploratory activities. For instance, SenseMap~\cite{suh2023sensecape} provides graph-based visual representations of users' browsing histories and the knowledge acquired during web-based sensemaking. Similarly, \citet{gratzl2016visual} explored recording users' exploratory interactions with a provenance graph during data analysis, allowing users to revisit previous analysis states and extend their explorations. Beyond capturing exploration histories, Sensescape~\cite{suh2023sensecape} supports non-linear exploration and organization of ideas through a spatial interface that accommodates multiple levels of abstraction. Related research has also explored sensemaking in the physical environment. Specifically, \citet{kim2025spatio} proposed structuring recorded video experiences into semantic events, key frames, and playback operations, enabling interactive and adaptive replay of past experiences. InsightAR~\cite{wu2026insightar} further supports the analysis of recorded physical activities by organizing and summarizing egocentric videos in a structured interface. Addressing the non-linear nature of complex workflows, \citet{yuan2026dependency} proposed a dependency-aware AR guidance that represents non-linear task workflows and identifies actionable steps based on their dependencies.

Collectively, these systems demonstrate the value of external representations for capturing, organizing, and interpreting exploratory activities. However, supporting sensemaking during ongoing physical problem solving presents additional challenge, as users must continuously interpret changes in their environment, evaluate the outcomes of their actions, and revise their understanding of the problem. This motivates tools that capture not only users' actions but also the surrounding environmental context and resulting state changes, enabling users to revisit observations, trace the progression of their investigations, and develop a more coherent understanding of the problem at hand.

\section{Formative Study}

We conducted a formative study using KeeBox~\cite{keebox}, a sequential discovery puzzle, to examine how individuals approach embodied problem-solving tasks with unknown solutions and explore the types of assistance they seek during the problem-solving process. The study involved seven participants (M=2, F=5, NB=0) aged 19 to 26 years (M=23, SD=2) in a group session with three KeeBox puzzles of different difficulty levels. 

Participants were first briefed by the study facilitator on the purpose of the study and provided with background information about the KeeBox puzzles. They were instructed to open the boxes without using force or causing damage and retrieve the hidden tokens inside. Participants were free to determine their own problem-solving approaches without following a prescribed procedure. They self-organized into three configurations: a group of three collaboratively solving the blue puzzle (difficulty 1/5), a group of three taking turns solving the purple puzzle (difficulty 2/5), and one participant independently working on the orange puzzle (difficulty 3/5). Following a 20-minute problem-solving session, all participants took part in a group semi-structured interview focused on their problem-solving strategies, challenges encountered, and desired forms of assistance.

We complemented the formative study with a literature analysis to contextualize our observations and discussions with participants, which led to the following design opportunities (DOs).

\paragraph{DO1. Unproductive Action Loops}
The cognitive load of embodied problem-solving taxes working memory, potentially causing users to lose track of recently attempted actions that yielded no progress. This leads to persistent repetition of unproductive actions, wasting cognitive and motor resources while inducing fatigue and stalling task progress.
This issue was observed in the formative study when participants repeated the same actions while brute-forcing a solution. Problem-solving literature offers a theoretical basis for this behavior in bounded working memory~\cite{baddeley1974psychology} and impaired metacognitive monitoring under
cognitive load~\cite{nelson1990metamemory,flavell1979metacognition}.

\paragraph{DO2. Strategy Tunneling}
When problem-solvers invest significant effort in a particular solution approach, they may become cognitively fixated on that strategy even when it proves ineffective, preventing them from considering alternatives that might be more productive. This phenomenon, known as the \textit{Einstellung} effect~\cite{luchins1942mechanization}, is rooted in the difficulty of reframing an initial mental representation, even when it does not lead to a successful solution~\cite{ohlsson1992information}.

In embodied problem-solving contexts, physical commitment to a configuration, such as arranging components in a specific way, can reinforce this functional fixedness~\cite{bilalic2008einstellung}. The invested motor effort and visible arrangement make it psychologically difficult to abandon the approach, as the initial schema blocks consideration of alternatives~\cite{bilalic2008einstellung}. Overcoming such fixation requires explicit support for strategy reflection and re-representation of the problem~\cite{ohlsson1992information,zhang1997nature}.

\paragraph{DO3. Track Problem-Solving Knowledge} Problem-solvers engaging with embodied problems often discover important constraints, relationships, and partial solutions, but might fail to retain or leverage this information as the task progresses. The cognitive demands of manipulating physical objects and exploring multiple solution paths can cause critical insights to fade from working memory before they can be integrated into the final solution~\cite{baddeley1974psychology,sweller1988cognitive}.

Embodied problem solving leaves few persistent traces of intermediate steps, unlike other forms of problem solving where intermediate results and progress can be recorded, such as scratch work when solving mathematical problems. Participants said maintaining such records during embodied problem solving would help, especially for longer problems, as external representations would capture knowledge distributed across internal and external states~\cite{zhang1994representations}.

\paragraph{DO4. Collaborative Problem Solving}
Collaborative problem solving can generate novel insights by integrating diverse perspectives, enabling emergent ideas that would not arise from individual cognition alone~\cite{nelson2013collaborative}. Participants who collaborated with peers explicitly mentioned that having multiple perspectives on the problem was helpful.

However, for embodied tasks, coordination challenges such as keeping all members focused on the same problem elements at critical moments, tracking what has been established versus revised, and establishing co-reference can limit the effectiveness of collaboration~\cite{barron2000achieving}. Therefore, productive collaborative problem solving requires tools that support a joint problem space by providing shared language, situation, and activity~\cite{roschelle1995construction}.

\begin{figure*}[ht!]
    \centering
    \includegraphics[width=.99\linewidth]{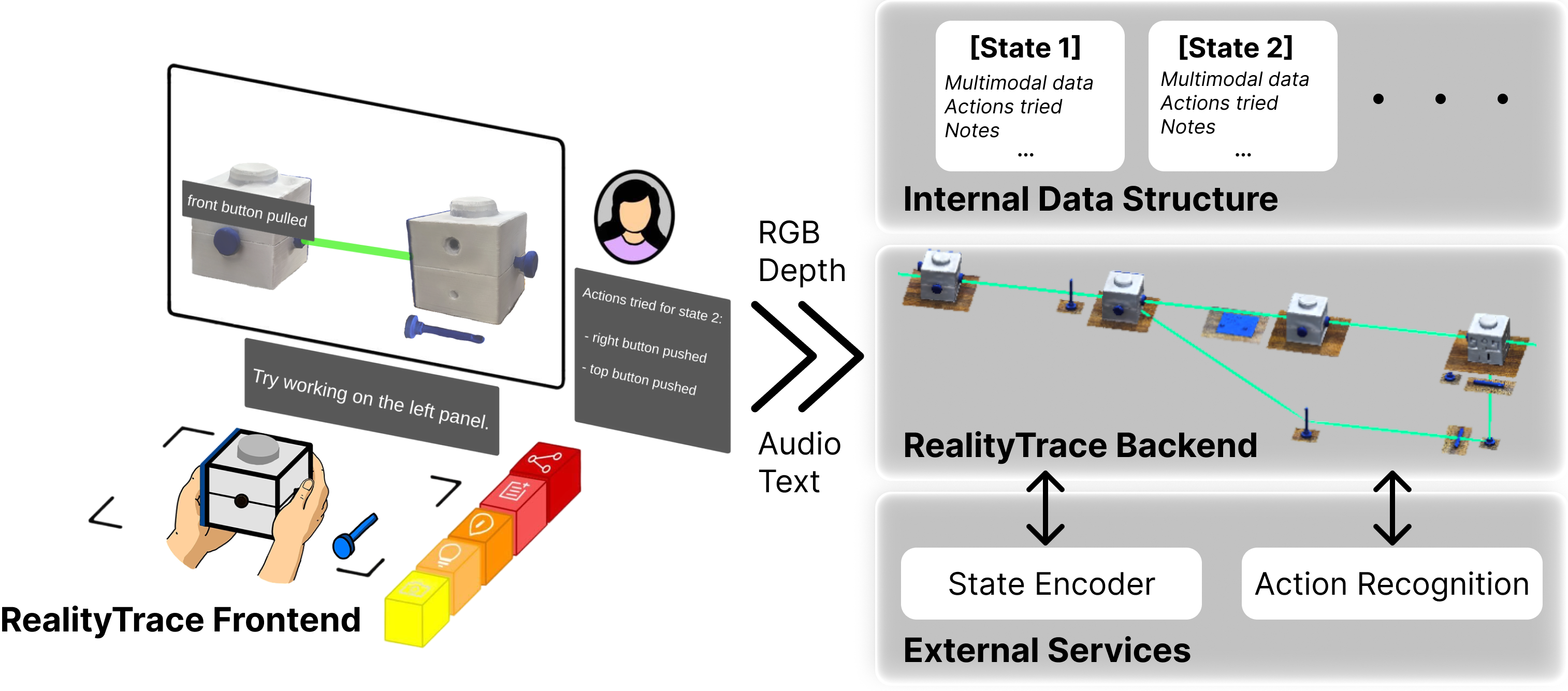}
    \caption{System architecture overview highlighting frontend-backend data exchanges and backend interactions for constructing the graph's internal states. The frontend captures and sends RGB image, depth image, audio, and text information to the backend, which interacts with external services (state encoder and action recognition) to build the graph representation of the problem space. The graph representation consists of multiple states connected by edges, where each state stores its multimodal data, actions tried, user notes, and more.}
    \Description{The figure presents the system architecture overview of RealityTrace, highlighting three main components (RealityTrace frontend, RealityTrace backend, and external services). The frontend captures and sends RGB image, depth image, audio, and text information to the backend, which interacts with external services (state encoder and action recognition) to build the graph representation of the problem space. The graph representation consists of multiple states connected by edges, where each state stores its 3D construction scan, user notes, actions tried, and more.}
    \label{fig:system}
\end{figure*}

\section{RealityTrace}

We designed RealityTrace to address the design opportunities identified in the formative study. In this section, we first describe its key features and then provide an overview of the system implementation.

\subsection{Features}

RealityTrace incorporates five key features that collectively address the four design opportunities identified in the formative study.

\begin{description}
    \item[F1. Action History] addresses \textit{Unproductive Action Loops (DO1)} by displaying a list of attempted actions in the current problem state as a text box in the user's field-of-view, as shown in~\cref{fig:teaser} (F1).
    
    \item[F2. Action Suggestion] addresses \textit{Strategy Tunneling (DO2)} by using the action history (F1) to suggest unexplored actions, encouraging users to consider alternative approaches rather than remaining fixated on an unproductive strategy. Suggestions can be triggered proactively at predefined time intervals or on demand by the user through a button click. Suggestions are displayed as a text box and as synthesized speech, as shown in~\cref{fig:teaser} (F2). Suggested actions aim to mitigate the \textit{Einstellung} effect~\cite{luchins1942mechanization, bilalic2008einstellung}, though it may not necessarily result in immediate progress.

    \item[F3. Problem-Solving Trace] addresses \textit{Track Problem-Solving Knowledge (DO3)} by creating a persistent external representation of the problem-solving process. When requested by the user, the system captures RGB frames and invokes the state encoding service to generate a 3D reconstruction of the problem state that is added as a node in a graph visualization. Edges between nodes represent the pragmatic actions that transform one state into another, creating a visual history of explored solution paths, as shown in~\cref{fig:teaser} (F3.a). The trace can be manipulated in augmented reality, allowing users to explore different perspectives and review previous states, as shown in~\cref{fig:teaser} (F3.b). This externalized trace offloads working memory demands and enables users to build on past discoveries rather than rediscovering the same insights repeatedly. 

    \item[F4. Spatial Notes] address \textit{Track Problem-Solving Knowledge (DO3)} by enabling users to attach persistent notes to specific problem states and physical workspace locations. Users can annotate both nodes in the trace (F3) and objects or their physical environment using text or speech input. These annotations remain anchored to their spatial context, making them retrievable when users return to similar problem configurations or physical arrangements, as shown in~\cref{fig:teaser} (F4). For example, a user might annotate a graph node with ``blocks center piece access'' or mark a workspace location with ``check alignment here.'' This spatial indexing supports memory retrieval by placing information where it is most contextually relevant.

    \item[F5. Shared Trace] addresses \textit{Collaborative Problem Solving (DO4)} by enabling multiple users to share and merge their problem-solving trace graphs, as shown in~\cref{fig:teaser} (F5). When users work on the same problem independently, their individual traces, which capture different explored paths, attempted configurations, and spatial annotations, can be combined into a unified collaborative representation. This merged graph reveals the union of all explored solution paths, allowing users to benefit from discoveries made by others without having to retrace those steps themselves. This distributed approach enables a divide-and-conquer strategy where the collective exploration of the problem space exceeds what any individual could achieve alone.

\end{description}

\subsection{Implementation}

The system architecture comprises a frontend that renders the RealityTrace interface and captures hand poses, head poses, and RGB images, and a Python backend that receives the captured data via WebSockets. The system is implemented using XARP, an open-source python toolkit for XR development~\cite{caetano2026xarp}. \Cref{fig:system} illustrates RealityTrace's modular architecture, which decouples external services from the core interface, enabling customization and allowing individual modules to be replaced as more advanced techniques become available. 

RealityTrace integrates two external services: (1) an \textit{action recognition} service that identifies user actions, and (2) a \textit{state encoder} service that generates multimodal representations of each graph node, including images and 3D reconstructions. As our primary contribution is the XR interface, we focus on demonstrating a functional integration of these services rather than optimizing their individual capabilities. The following sections describe our preliminary implementations, whose performance may vary depending on the underlying architectures, methods, and vision-language models used. Further optimization of these services is beyond the scope of this work.

\subsubsection{Action Recognition}

We implement the action recognition service as an agentic system for egocentric video understanding, using LangGraph to orchestrate the workflow and Qwen3.8-27B~\cite{qwen38_27b} as the vision-language model backbone. The system processes headset video frames a configurable sampling rate of 1 frame per second by default. This rate can be adjusted based on available computational resources and real-time processing requirements to balance recognition accuracy and latency. The agent analyzes consecutive sampled frames to identify visible objects and user's hands, describe their states and relationships, and infer user actions. To maintain object identity across frames, it attempts to associate detected objects with existing object IDs, assign new IDs to previously unseen objects, and leaves ambiguous associations unresolved when necessary. The agent stores states and relationships as timestamped facts and actions as separate events. For example, it can record that a puzzle piece is resting on the table, then log a pickup event associating the user's hand with that piece. Facts and events retain confidence estimates and references to their source frames. When an action cannot be reliably inferred, the agent can examine additional nearby frames to resolve the ambiguity.

\subsubsection{Action Suggestion}

\textit{F1. Action History} also serves as the basis for \textit{F2. Action Suggestion}. Since the full action space and the solution state are initially unknown and progressively discovered during problem solving, suggestions must rely solely on information collected throughout the task. We therefore adopt a zero-prior-knowledge approach that builds a discovered action space from previously observed actions (e.g., knob A can be pressed and knob B can be turned) and compares it with actions attempted in the current problem state to identify unexplored alternatives.

\subsubsection{State Encoding}

The state encoding service is implemented using Meta SAM 3D Objects~\cite{sam3dteam2025sam3d3dfyimages}, which generates a 3D model of an object from an input image. This model was selected for its computational efficiency, ability to automatically segment object of interest from their surrounding environment, and capability to produce 3D models without requiring users to actively rotate objects to capture multiple views.

\section{Usability Study}

To assess RealityTrace's usability and usefulness for embodied problem solving, we conducted a study with eight participants who used RealityTrace to solve a Keebox puzzle. To ensure consistent conditions across participants without interference from the non-deterministic behavior of the state encoding and action recognition services, we replaced these components with Wizard-of-Oz simulations~\cite{kelley1984interactive}.

\subsection{Participants}
We recruited eight participants (M=4, F=4, NB=0), aged 21 to 27 years (M=23, SD=2), from our institution. All participants had normal or corrected-to-normal vision and provided informed consent before participation. Seven participants had prior experience with extended reality. Participants reported prior experience with various puzzle games, including Rubik's Cube ($N=4$), Rush Hour ($N=1$), jigsaw puzzles ($N=4$), Sudoku ($N=6$), Tetris ($N=4$), and tower of Hanoi ($N=2$). Each participant was compensated for their time with an Amazon gift card of \$10. This study was approved by our local IRB (\textit{anonymized protocol}).

\subsection{Apparatus}
Participants used the RealityTrace interface on a Meta Quest 3. The study used KeeBox Blue (level 1/5) for the experimental task and KeeBox Purple (level 2/5) for the task familiarization phase. The system backend was connected to a web interface operated by the Wizard, who watched the user's egocentric video stream from the headset on a separate screen.

We simulated the state encoding service by mapping the puzzle solution as a graph with alternative paths, as illustrated in \cref{fig:state_machine}. For each puzzle state, we pre-scanned and generated a 3D reconstruction using Polycam~\cite{polycam}. These reconstructions were reused across all participants to ensure consistency and fidelity throughout the study.

To simulate the action recognition service, we first curated a list of possible actions for each state as action-component pairs (e.g., flipping the box, pressing a button), including both actions that advance the puzzle toward its solution and those that do not contribute to progress. During the study, the wizard monitored participants' actions through the live video feed and identified completed actions by matching their observed behavior against the pre-curated list. The wizard then marked the corresponding actions as completed through the web interface, which automatically updated the action history displayed to the participants (F1). 

The action suggestion feature (F2) was automatically triggered every 30 seconds. Each suggestion was picked by excluding actions already performed in the current puzzle state from the pre-curated list of possible actions and randomly selecting one from the remaining candidates.

Participants used the capture feature to construct the problem-solving trace. Each time a participant activated the feature, an audio cue notified the wizard, who then updated the problem-solving trace by loading the corresponding puzzle state based on the participant's progress.

\subsection{Procedure}

After providing informed consent, participants completed a demographic questionnaire and watched a two-minute video introducing the KeeBox puzzle. Participants then entered a task familiarization phase, during which they were given five minutes to explore KeeBox Purple, familiarize themselves with the puzzle, and make as much progress as possible. To support naturalistic problem solving, participants were allowed to use common tools, such as notebooks, sketches, and phone cameras. Immediately afterward, we conducted a five-minute semi-structured interview to understand participants' challenges and problem-solving strategies.

Following the familiarization phase, participants watched a three-minute tutorial introducing RealityTrace and its features. They were then given KeeBox Blue, which requires similar types of actions but features a different configuration and solution to minimize learning transfer from the familiarization phase.

\begin{figure*}[ht!]
    \centering
    \includegraphics[width=.99\linewidth]{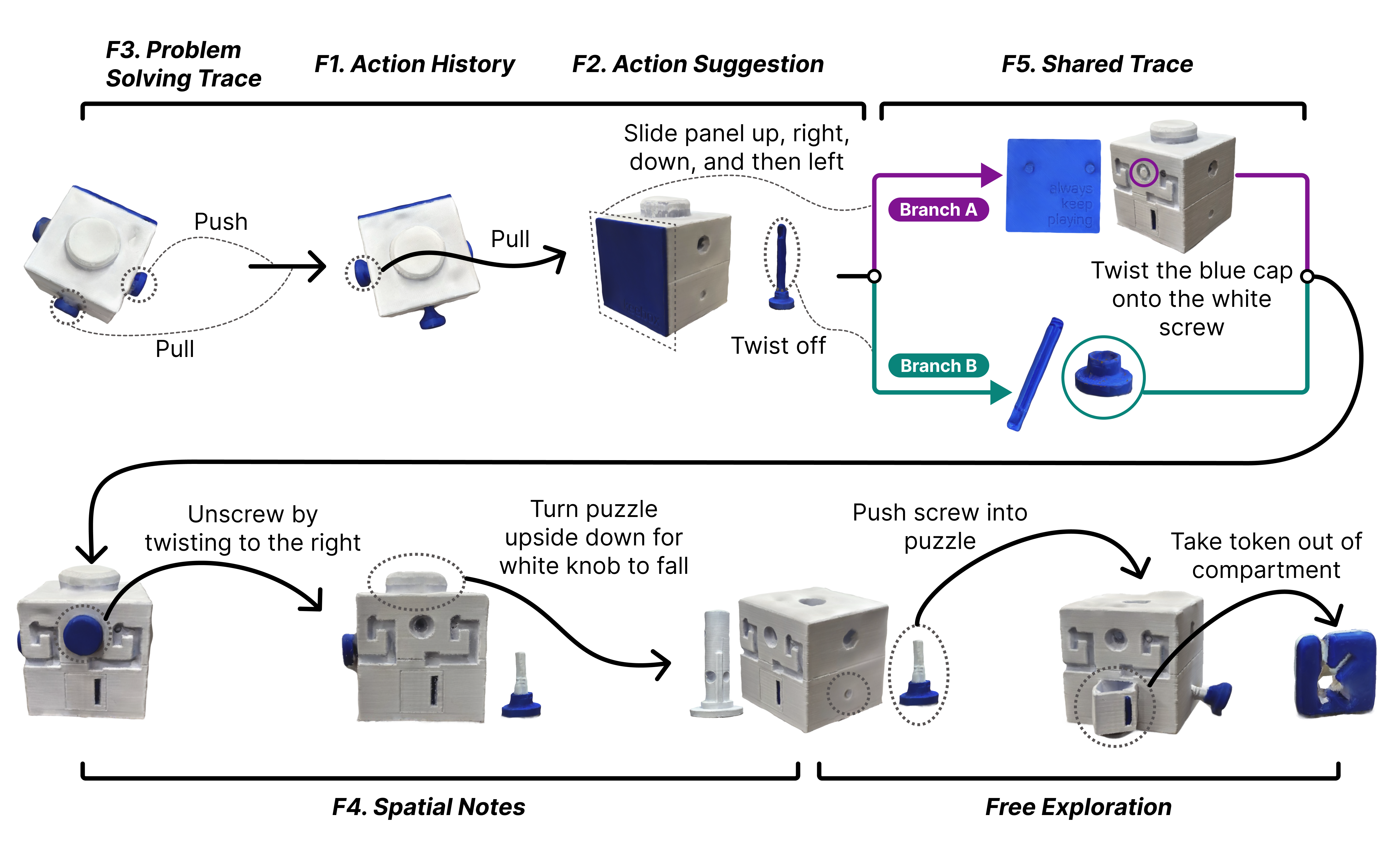}
    \caption{Solution trace annotated with transition actions leading to each subsequent state. Feature labels above and below the trace indicate when each feature was introduced during the solution process, followed by a free exploration phase in which users could freely interact with all system features.}
    \Description{The figure presents a step-by-step solution trace of a Keebox puzzle, illustrating how specific transition actions lead to successive states. Starting from initial manipulations (push, pull, slide, and twist), the diagram tracks the sequence of interactions required to progress through the puzzle. Branching paths (A and B) indicate alternative intermediate actions that later converge on a shared step, where a blue cap is twisted onto the puzzle box. Subsequent actions include unscrewing, inverting the puzzle to release an internal component, reinserting the screw, and extracting a hidden token. Annotations on the right categorize the information available to participants at each stage. The first stage includes problem-solving trace (F3), action history (F1), zero-prior-knowledge suggestion (F3). The second stage includes a shared trace (F5). The third stage includes spatial notes (F4) The final stage is free exploration.}
    \label{fig:state_machine}
\end{figure*}

Participants then began the main study phase, during which the experimenter guided them through a predefined procedure in the following order: 

\begin{enumerate}
    \item Participants were instructed to scan the initial state of the puzzle to initialize the problem-solving trace visualization (F3). The experimenter then provided verbal instructions on how to pan and rotate the trace for closer inspection.

    \item Participants were then instructed to begin solving the puzzle and were reminded to check the action history (F1) displayed on the interface as they performed actions.

    \item When participants heard an audio narration of an action suggestion (F2), the experimenter reminded them that it was a proactive action suggestion from the system. Participants were also instructed to manually trigger the action suggestion feature (F2) to explore its on-demand functionality.

    \item When participants reached one of the branching paths (see \Cref{fig:state_machine}), they were instructed to clicked the share button to introduce a virtual collaborator (F5), who then updated the problem-solving trace with the other branch.

    \item The experimenter guided participants through the spatial notes feature (F4), instructing them to create one world-anchored note and one model-anchored note.
\end{enumerate}

The experimenter-facilitated phase lasted 20-25 minutes, followed by a 10-minute period during which participants could freely use the system to solve the puzzle without further guidance from the experimenter. Two of the eight participants successfully solved the puzzle. \Cref{fig:state_machine} illustrate the solution trace and indicates the steps at which each feature was introduced.

Finally, we conducted a semi-structured interview about feature preferences and scenarios where participants believed the system would be helpful. Participants completed a SUS survey~\cite{bangor2009determining}. Interviews were recorded and automatically transcribed.

We analyzed semi-structured interview transcripts and researcher notes using reflexive thematic analysis~\cite{braun2006thematic, braun2025report}. Two researchers began by familiarizing themselves with the corpus, then independently coded the data and grouped codes into initial themes. Next, they met for two hours to compare, discuss, and refine their themes. The final themes were named as presented in section~\ref{sec:findings}.

\section{Results}
\label{sec:findings}

The overall SUS score for RealityTrace was 65.94, placing it in the marginal high range~\cite{bangor2009determining}. We now present the themes developed from the data collected in our usability study.

\subsection{When Structure Pays Off} 

During the familiarization stage, participants were instructed to make as much progress as possible on the puzzle, with paper for note-taking and a phone camera available for use. However, none of them spontaneously used these resources to document their problem-solving process or help them in finding the solution. Participants generally did not perceive such documentation as necessary for the task. For instance, P1 explained, ``I thought recall doesn't help me,'' and P3 stated, ``I think I don't need to memorize anything.''

In contrast, when using RealityTrace, participants engaged with its structured representations and reported benefits for problem solving. Beyond documenting processes, RealityTrace enabled participants to receive action suggestions (F2), inspect recorded states and actions through an interactive visualization (F3), and collaborate through shared problem-solving histories (F5). This contrast suggests that the perceived usefulness of external representations depended not merely on their availability, but also on the additional capabilities they enabled.

The value of these representations was not limited to immediate problem solving. Participants also recognized opportunities to revisit and reflect on their problem-solving processes. P7 noted that having their actions recorded ``made me think oh am I doing this right or am I doing this wrong,'' while also making them aware of repeated actions. Similarly, P2 appreciated that the branching structure allowed them to ``backtrack to get to the point where you can start branch B,'' and anticipated revisiting the documentation when encountering variations of previously solved problems. P2 also envisioned using the system to ``share my train of thought and the steps that I took to solving this problem.'' These responses suggest that preserving the problem-solving process, rather than only its outcome, could support self-evaluation, exploration of alternative approaches, and future knowledge reuse.

However, participants did not perceive all forms of external representations as equally valuable. Several participants made limited use of the spatial note-taking features, citing both interaction effort and redundancy with the system's existing documentation. Creating a spatial note required additional interactions, including button click, anchor placement, and typing through the Meta Quest system keyboard. P4 described typing as ``kind of exhausting'' and explained that they did not find taking notes particularly helpful because the system already tracked puzzle states and actions. This suggests that the value of manually creating additional representations may be limited when similar information is already captured by the system, particularly when the added interaction effort is substantial.

Participants' experiences further indicated that the timing and degree of automation influenced how assistance was perceived. Although action suggestions (F2) were generally viewed positively, preferences for automatic triggering varied. P2 preferred receiving suggestions on demand, explaining, ``I would like to use my own brain first before it gets these suggestions.'' In contrast, P4 appreciated receiving suggestions automatically because they sometimes became focused on other aspects of the puzzle and overlooked available next steps. Similar considerations emerged for state capture, where some participants valued the possibility of automatic scanning to reduce manual effort, while also wanting to retain manual controls to address potential errors. For instance, P4 suggested retaining the manual option ``so I can correct it.'' Together, these responses indicate that the perceived usefulness of automated support depends not only on reducing interaction effort, but also on how and when assistance is initiated~\cite{luo2026users, chen2025need} and on users' ability to intervene when necessary~\cite{horvitz1999principles, amershi2019guidelines}.

\subsection{Participants Envisioned Applications}
Based on reflection and prompting, participants identified several additional applications for RealityTrace beyond puzzles.

One prominent domain is \textit{device repair and troubleshooting}, which often involves iterative testing, disassembly, and reassembly, requiring users to keep track of intermediate states and actions. Participants envisioned using RealityTrace to document these processes through its problem-solving trace (F5) and storing rich multimodal data that facilitates efficient lookup, verification, and documentation of results. For example, P4 described replacing a laptop fan, where tracking removed screws and disconnected wires was essential for correctly reassembling the device. P2 further envisioned remote troubleshooting with an expert, emphasizing the value of revisiting previous states: ``Knowing that I can backtrack to previous state, because it's visually available to me is valuable.'' These scenarios highlight RealityTrace's potential to support traceable repair workflows, reduce reliance on memory, and facilitate collaborative troubleshooting.

Another promising domain is \textit{scientific workflows}, which typically involve structured, sequential processes that benefit from systematic recordings. P7 envisioned using RealityTrace in chemistry laboratories to ``record the steps for the experiment,'' enabling researchers to revisit previous states and document experimental procedures. P7 also suggested that branching traces could capture alternative experimental paths and outcomes. These possibilities highlight RealityTrace's potential to support experimental documentation, comparisons of alternative approaches, and reproducibility.

Additionally, P5, drawing on experience with software version control, compared RealityTrace to Git, suggesting it can record workflows, ``ask your friends for their advice... and you can merge them into your own branch.'' This points to RealityTrace's potential as a collaborative hub, where problem-solving process can be distributed, shared, and integrated among multiple contributors.

\begin{figure*}[ht!]
    \centering
    \includegraphics[width=.99\linewidth]{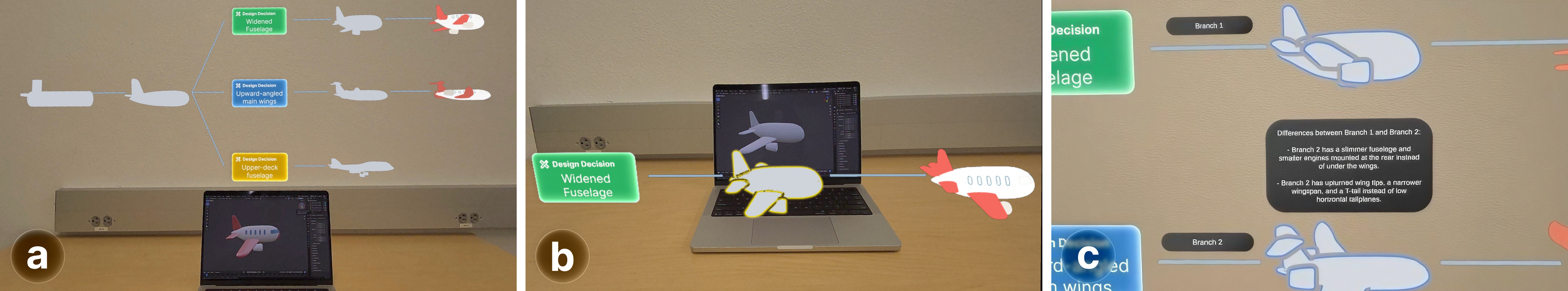}
    \caption{Blender with RealityTrace plugin: (a) RealityTrace maintains a visual history of all design explorations, with annotated nodes capturing the design decision underlying each iteration. (b) RealityTrace allows users to revisit previous design states by restoring the corresponding model in Blender, allowing them to resume editing and explore alternative design directions. (c) RealityTrace inspects and compares two user-selected models and presents their key differences in a text box, allowing users to examine the evolution of their designs or identify distinctions between alternative design branches. \textit{The 3D aircraft models were created using Codex and Blender to demonstrate the design iteration process.}}
    \Description{The figure presents three screenshots of how RealityTrace can be used as a plugin for Blender. The first screenshot (a) shows how RealityTrace shows the branching design iterations made by the user, with annotated nodes highlighting the design decision for each iteration. The second screenshot (b) shows how RealityTrace can restore the corresponding model in Blender when a user selects that model in the graph trace. The third screenshot (c) shows how RealityTrace can inspect and outline the key differences between two user-selected models.}
    \label{fig:blender_application}
\end{figure*}

\begin{figure*}[ht!]
    \centering
    \includegraphics[width=.99\linewidth]{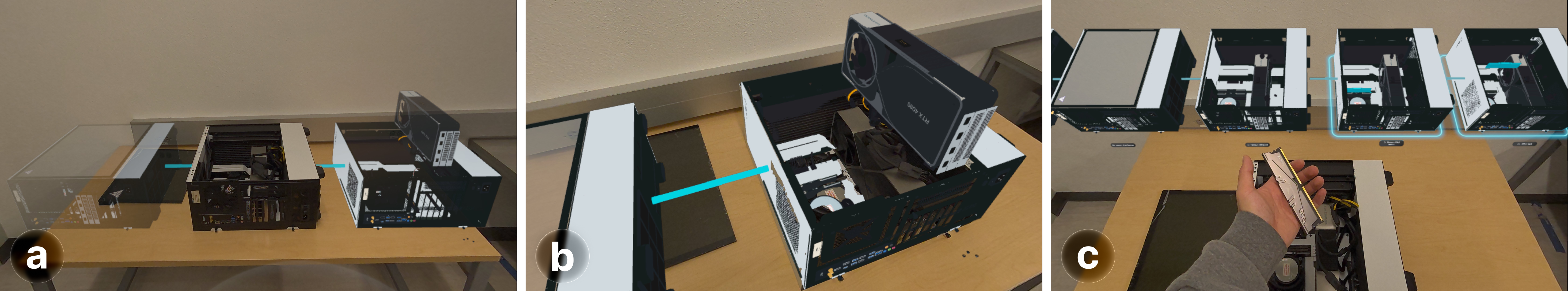}
    \caption{RAM replacement with RealityTrace: (a) The spatially contextualized view of their progress helps users identify their current position within the overall problem-solving process and determine subsequent steps. (b) Users can overlay their desired target state onto the physical device, enabling direct visual comparison between the current and target configurations. (c) RealityTrace retrieves relevant steps associated with specific physical components, supporting object-based queries and context-aware navigation through the problem-solving history. \textit{The 3D models illustrating the different stages of PC assembly and disassembly were created using Codex and Blender, based on actual photos taken on the PC.}}
    \Description{The figure presents three screenshots of how RealityTrace can be used for computer repair, specifically RAM replacement. The first screenshot (a) shows how the contextualized view of the user's progress can help them identify their specific position within the overall problem-solving process. Their physical state becomes part of the actual trace. The second screenshot (b) shows how users can overlay the desired target state on top of their physical state for direct visual comparison. The third screenshot (c) shows how RealityTrace can retrieve the relevant steps that are associated with the physical component the user interacts with. This allows for more flexible navigation through the problem-solving history.}
    \label{fig:repair_application}
\end{figure*}

\section{Applications Beyond Puzzles}

To illustrate the generalizability of RealityTrace across a broader range of problem-solving tasks, we present the following walkthroughs of representative usage scenarios. The accompanying images include both AI-assisted visual mockups and captures from an actual headset, with AI tools also used to support the development of these examples (e.g., 3D model generation). These examples highlight RealityTrace's potential to augment human capabilities across diverse applications.

\subsection{Iterative Design Process}

The iterative design and prototyping process can be viewed as a form of problem solving, in which the designers progressively refine design artifacts, explore and evaluate alternative solutions, and iterate toward achieving a desired design goal~\cite{camburn2017design}. This process can be captured as a problem-solving trace, which not only serves as a structured record of the design process but also facilitates systematic versioning, comparison, and backtracking across design iterations.

Consider a 3D modeling workflow in which RealityTrace has been integrated into design software such as Blender. As designers iteratively refine their 3D models, they can mark meaningful design milestones as states in RealityTrace, with the current 3D model automatically retrieved from Blender and stored as part of the corresponding node. When adding a new state, designers can also annotate the edges connecting successive states to document the design decisions and rationale behind each iteration. Designers can then navigate the resulting trace to review the evolution of their designs, compare alternative versions, or return to an earlier state, which automatically restores the corresponding 3D model in Blender for further editing and exploration.

Such integration transforms the design history into an interactive and navigable representation of the design process, allowing designers to revisit earlier decisions, explore alternative design directions, and branch from previous iterations without losing their progress. By preserving both design artifacts and the reasoning behind their evolution, RealityTrace can reduce the overhead of manual documentation and version management while supporting more informed and flexible design exploration.

\subsection{Computer Repair}

Inspired by the computer repair use case suggested by participants in the usability study, we demonstrate how RealityTrace can support computer RAM replacement, a task involving the disassembly and reassembly of a computer.

When disassembling an unfamiliar computer, users may need to incrementally explore its internal structure and determine how to access the target component, without knowing the required sequence of actions in advance. RealityTrace captures this exploratory problem-solving process as users interact with the device. During reassembly, users can revisit their captured traces to retrace and reverse their previous actions, reducing the need to remember the specific steps taken during disassembly.

With the 3D representation of captured states, users can visualize the problem-solving trace in situ alongside the physical computer. By displaying the preceding and subsequent steps directly next to the device, RealityTrace provides users with a spatially contextualized view of their progress, helping them understand their current position within the overall process. Additionally, users can overlay a desired target state onto the physical device, enabling them to visually compare its current configuration with the target configuration represented in the trace. 

Beyond visualizing the overall disassembly and reassembly process, RealityTrace can also enable object-based queries, allowing users to retrieve relevant steps through interactions with physical components. For instance, when a user picks up a disassembled computer component, RealityTrace can identify the corresponding steps in the recorded trace by comparing the component's 3D geometry against the captured states. This enables users to quickly locate steps associated with a specific component without manually searching through the entire trace, supporting more flexible and context-aware navigation of their problem-solving history.

\subsection{World-Scaled Usage}

Beyond tracking and documenting object-scale problem-solving processes, RealityTrace can also support larger-scale activities involving the surrounding environment and human body. Rock climbing, for example, presents an exploratory problem-solving task in which climbers must determine how to navigate a route with multiple possible movement sequences. Climbers typically approach this through repeated attempts, experimenting with different techniques and progressively refining their strategies.

RealityTrace can support this iterative problem-solving process by capturing climber's attempts, including their movement sequences and techniques. During subsequent attempts, climbers can revisit and reuse previously recorded trace to recall successful movements, identify ineffective approaches, and inform their next actions. In this way, RealityTrace enables climbers to build upon previous experiences and progressively develop a solution to the climbing route.

RealityTrace captures these climbing experience through representing the climber's body and its interactions with the surrounding environment. At a fine-grained level, precise hand positioning and grip techniques are critical for maintaining stability and progressing along a route. RealityTrace enables climbers to capture their hand poses on individual holds, preserving these details for reference in subsequent attempts. When a hold supports multiple viable grip techniques, users can capture alternative hand poses alongside the reconstructed geometry of the surrounding climbing wall, allowing them to visualize and compare different approaches within their spatial context. At a broader level, RealityTrace can leverage external sensors, such as ground-based camera tracking system, to capture full-body postures and movement sequences, which can then be visualized in situ on the climbing wall. Together, these representations enable climbers to revisit their previous attempts at different levels of detail, from individual grip techniques to overall movement strategies, and progressively refine their approaches across repeated attempts.

\begin{figure*}[ht!]
    \centering
    \includegraphics[width=.99\linewidth]{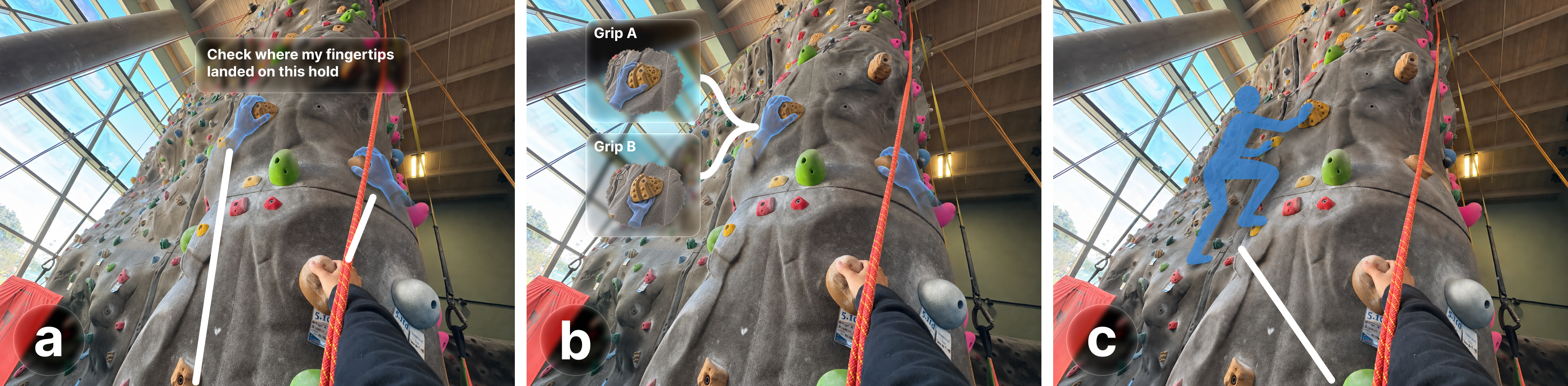}
    \caption{Rock climbing with RealityTrace: (a) Users can place spatial annotations to document unsuccessful attempts, reflect on previous climbs, and iteratively refine their strategies across subsequent attempts. (b) For holds that support multiple grip techniques, users can capture and compare alternative hand poses to explore different approaches. (c) With external sensors, RealityTrace can capture full-body movement sequences for visual guidance on subsequent climbs. \textit{The background images for the egocentric views were captured at a climbing gym, while the virtual hands, alternative grips, and virtual body were generated using ChatGPT Image.}}
    \Description{The figure presents three pictures of how RealityTrace can be used for rock climbing. The first picture (a) shows how users can place spatial annotations on the climbing walls to document previous unsuccessful attempts and reflect on their climbing techniques so they can subsequently refine their next steps. The second picture (b) shows how users can capture alternative hand poses with holds that support multiple grip techniques. The third picture (c) shows how RealityTrace can be used with external sensors to capture the user's full-body movements as a more comprehensive visual guidance tool for subsequent climbs.
}
    \label{fig:climbing_application}
\end{figure*}

\section{Design Reflection}

Building on our system design, implementation, and example use cases, we reflect on key design considerations and broader insights gained from developing RealityTrace, particularly regarding the role of assistive technologies in supporting human problem solving.

\subsection{Bimanual Input and Visual Outputs Overhead}

Although RealityTrace supports controller-free hand interaction, users must split actions between task manipulation and system controls. This may divert attention from the task and disrupt the natural flow of physical actions. Grasp user interfaces could reduce this friction by enabling system interaction while users hold and manipulate objects~\cite{aponte2024grav, sharma2024graspui, caetano2025graspr}. Similarly, embedded visualizations could reduce attention fragmentation by presenting information directly on objects~\cite{willett2016embedded}.

\subsection{Sharing Representations with AI Agents}
RealityTrace creates structured representations of embodied problem solving that are both human- and agent-readable, enabling human-AI collaboration even without physical embodiment of the agent. Through callable tools, AI agents can use the same primitives as users, such as spatial manipulations and deictic references, enabling co-reference. Systems for embodied problem-solving support should preserve user agency over what is captured and shared, preserving autonomy and privacy. Future iterations should consider structured representations of sound, pressure, and weight to capture richer context.

\subsection{From Process Documentation to Reflective Support}

While our findings suggest that preserving problem-solving histories can create opportunities for reflection, they also reveal a distinction between documenting a process and actively supporting reflection regarding a process. Our system primarily captures intermediate states and actions to assist users during problem solving. However, participants' responses suggest that these representations may also encourage users to reconsider their actions, revisit alternative approaches, and apply previous experiences to future tasks. This points to an opportunity for designing systems that support not only the reconstruction of \textit{what} happened, but also reflection on \textit{why} a particular decision was taken and how they might be improved. For example, future systems could help users analyze repeated actions, revisit key decision points, or compare alternative solution paths.

\subsection{Goal-Aligned Assistance}

The puzzle task used in the study led some participants to express a preference against receiving hints, as such assistance distracts from the intrinsic enjoyment of problem solving. This observation raises a broader question about the goals of problem-solving assistance: whether systems should prioritize immediate task outcomes, particularly in time-sensitive scenarios, foster a deeper understanding of the problem at hand, or support the long-term development of independent problem-solving skills. Rather than treating these objectives as mutually exclusive or universally prioritizing one over the other, this tension highlights an opportunity to design \textbf{goal-aligned assistance} that accommodates different priorities, balancing immediate task performance with deeper understanding and long-term cognitive development according to users' preferences and task contexts.

\section{Conclusion, Limitations \& Future Work}

While we evaluated user interaction with RealityTrace in a usability study, these findings are time-limited; how users experience the system over longer sessions and across other task scenarios remains an open area for future work. To isolate the interface and examine how the system supports problem solving with access to 3D reconstructed representations, we employed a Wizard-of-Oz approach to control the quality of system-generated suggestions and action tracking. However, the quality of these components is likely to influence user experience in practice. Future work should investigate how the quality of these features impacts user experience.

\section{Generative AI Use Disclosure}

AI tools were employed to support software development activities throughout the system development process, as well as the creation of visual assets for the illustrative examples, including visual mockups and 3D models. LLMs were used to suggest recommendations for improving text phrasing and clarity. Any suggestions provided by LLMs served only as input to our own decision-making.

\bibliographystyle{ACM-Reference-Format}
\bibliography{main}


\end{document}